# Voltage Control of Uni-directional Anisotropy in Ferromagnet-Multiferroic System

Sasikanth Manipatruni[1], Dmitri E. Nikonov[1], Chia-Ching Lin[1], Prasad Bhagwati[2], Yen Lin Huang[2], Anoop R. Damodaran[2], Zuhuang Chen[2], Ramamoorthy Ramesh[2] and Ian A. Young[1]

[1]Components Research, Intel Corp., Hillsboro, Oregon, 97124, USA

[2]Department of Materials Science and Engineering and Department of Physics, University of California, Berkeley, California 94720, USA

Demonstration of ultra-low energy switching mechanisms is an imperative for continued improvements in computing devices[1-3]. Ferroelectric (FE) and multiferroic (MF)[4,5] orders and their manipulation promises an ideal combination of state variables to reach atto-Joule range for logic and memory (i.e., ~ 30X lower switching energy than nanoelectronics)[6,7]. In $BiFeO_3$ the coupling between the antiferromagnetic (AFM) and FE orders is robust at room temperature[8, 9], scalable in voltage[10], stabilized by the FE order[4], and can be integrated into a fabrication process for a beyond-CMOS era[11,12]. The presence of the AFM order and a canted magnetic moment in this system[13] causes exchange interaction with a ferromagnet such as CoFe[14, 15] or LSMO[16]. However, what is needed/preferred for logic and memory is a magneto-electric mechanism that works analogous to an applied field, i.e a uni-directional anisotropy that is voltage modulated. While previous work has shown that exchange coupling (uniaxial anisotropy) can be controlled with an electric field, several puzzling issues remain. Perhaps the most intriguing among them is that the $BiFeO_3$-CoFe bilayer did not demonstrate any electrically controlled directional anisotropy, i.e., an exchange bias, which is a potential mechanism for $180^o$ magnetic reversal and is independent of switching kinetics[14]. Here, we present the evidence of electrical control of exchange bias of a laterally scaled spin valve that is exchange-coupled to $BiFeO_3$ at room temperature. We show that the exchange bias in this bilayer is thermally robust, electrically controlled and reversible. We anticipate that magneto-electricity at such scaled dimensions provides a powerful pathway for computing beyond the modern nanoelectronics transistors by enabling a new class of non-volatile, ultra-low energy computing elements.

The exponential scaling in computing device density and concurrent gains in the efficiency of computation relies on reducing the device-level switching energy[1-3]. Nanoelectronics has entered an exciting and yet challenging phase when the transistor dimensions have already been scaled to ~ 20 nm. Further device-level improvements are likely to involve new physical mechanisms and new state variables[18]. In this context, the discovery and development of novel devices suitable for computing has become imperative. Magneto-electric switching provides a new and potentially ultra-low energy switching mechanism, since the energy density for switching (E/A= $2.P_C.V_{fe}$, where $P_c$ is the polarization per unit area, $V_{fe}$ is the ferro-electric switching voltage) can be scaled to 1 µJ/cm$^2$, a significant leap in energy efficiency compared to modern CMOS (E/A= ½ $C.V^2$). Further, the multiferroic nature of the switching medium enables non-volatility (relying on the FE polarization *P* and the Anti-Ferromagnetic order *L*), directionality and signal regeneration. Multiferroic magneto-electrics can enable aJ class logic and memory [7, 19], where a voltage driven multiferroic switches the magnetization of a ferromagnet (Fig. 1).

The ability to switch a ferromagnet by generation of an effective magnetic field or unidirectional anisotropy at the interface enables a new class of devices for integrated circuits (example Fig1 A. A magneto-electric logic device[7] switched with voltage control of magnetism). The preferred magneto-electric switching method for device applications is the generation of electrically controlled exchange bias to switch a ferromagnet (FM) in direct contact with the multiferroic (MF). While spin-torque driven switching of a ferromagnet has been well studied showing the existence of both field-like and damping-like torques [28,29], magneto-electric switching torques on a ferromagnet due to a magneto-electric multiferroic have not been described. Electrically controlled exchange bias (EB): a) modifies the energy landscape of the magnet asymmetrically whereas exchange coupling (EC) produces a uniaxial change to the magnetization landscape (Fig 1 b, c); b) allows wider range/independent switching dynamics for the FM in a logic/memory device. *Ab initio* studies of the multiferroics [20-22] suggest that within a single ferroelectric domain, there exists the possibility of a directional magnetic coupling, i.e., the existence of an electrically switchable exchange bias, due to the presence of the canted magnetization [23,24,30-33]. The multiferroic can therefore impose an electric field controlled exchange bias onto a ferromagnet that is in contact with it; surprisingly this has not been experimentally reported so far. Prior work on BiFeO$_3$ (BFO), the only reliably high temperature magneto-electric multiferroic[14-16], has shown the ability to control the direction of magnetization in a ferromagnetic layer via exchange coupling to the multiferroic surface. However, electrical control of exchange bias with multiferroics has been limited to low temperature [25] with LSMO as the ferromagnet. Cr$_2$O$_3$ [26] exhibits a robust exchange bias but lacks the FE order limiting non-volatility and directionality for device applications.

In this work, we demonstrate that there is indeed a robust, electrically switchable exchange bias at the interface between the BFO layer and a dimensionally scaled CoFe/Cu/CoFe GMR spin valve. A key to the observation of the exchange bias, appears to be the reduction in the lateral dimensions of the GMR stack, with respect to the ferroelectric domains in the BFO layer. Furthermore, this exchange bias is reversibly modulated by a bipolar electric field, again confirming the directional coupling of magnetism and its control with an electric field. Thus electrically controllable exchange bias is a key enabler for a new class of logic and memory devices (including the magneto-electric spin orbit (MESO) device) for reducing the energy per transition of a solid state switch beyond the modern nanoelectronic transistors. It is poised to enable new class of non-volatile, ultra-low energy computing devices.

The description of the magneto-electric switching torques (from exchange bias and exchange coupling) in a magneto-electric switch is a critical first step for utilizing magneto-electricity. If a ferromagnet with magnetization **M$_{FM}$** is coupled to a multiferroic magneto-electric anti-ferromagnet (where **M$_1$, M$_2$** are the moments of the magnetic sub-lattices of the anti-ferromagnet), the coupling is most conveniently expressed via the uncompensated magnetization due to canting of the sub-lattice moments **M$_c$=M$_1$+M$_2$**, and the antiferromagnetic order **L=M₁-M₂**. The switching of the ferromagnet can be described by the Landau-Lifshitz-Gilbert [20] equation modified with magneto-electric coupling (ME-LLG) terms:

$$\frac{\partial m_{FM}}{\partial t} = -\gamma\mu_0 \left[ m_{FM} \times \vec{H}_{eff}\left(\vec{L}, \vec{M}_c\right) \right] + \alpha \left[ m_{FM} \times \frac{\partial m_{FM}}{\partial t} \right] \quad (1)$$

Where the effective field includes the contributions from the shape anisotropy, magneto-crystalline anisotropy, exchange coupling and exchange bias:

$$\vec{H}_{eff} = \vec{H}_{sh} + \vec{H}_{mcry} + \vec{H}_{ec} + \vec{H}_{eb},$$

$$\vec{H}_{ec}(\vec{L}) = H_{ec}\hat{l}(\hat{l} \cdot \hat{m}_c),$$

$$\vec{H}_{eb}(\vec{M}_c) = H_{eb}\hat{m}_c. \quad (2)$$

Here lower case letters with hats denote unit vectors along the corresponding upper case vectors. Furthermore $\vec{H}_{eb}(\vec{M}_{me}) \bullet \vec{H}_{eff}(\vec{L}) = 0$ since $\hat{L} \bullet \hat{m}_c = 0$ for equivalent sub-lattices. It is known that in BFO, vectors **P**, **L** and **M_c** undergo a kinetic two-step process in response to an electric field [14] (please see Figure 2, section 1 of supplementary). Figure 2 shows the trajectory of **L** and **M_c**, following the trajectory of the polarization **P**, where **P** under goes a non-Vanderbilt switching mechanism (where the **P** rotates avoiding a zero polarization state). Accordingly, **M_c** produces a reversible exchange bias when **P** is reversed via application of out of plane voltage. The torque arising due to coupling to anti-ferromagnetism ($\tau_L$, TAFM) and torque due to coupling to canted ferromagnetism **M_C** ($\tau_{Mc}$, TMC) affect the ferromagnet in a different way: $\tau_L$ produces a steady state anisotropy change (exchange coupling) and $\tau_{Mc}$ produces a steady state effective magnetic field (exchange bias). If $\tau_{Mc}$ is stronger than the combination of anisotropies, it can cause magnetization reversal. $\tau_{Mc}$ defines a switching trajectory for the FM, so it does not need to respond in a time scale close to FE time scale. In particular for (001) lattice orientation of the BFO film, $\tau_{Mc}$ can also exert an out of plane torque due to an out-of-plane projection of the canted magnetic moment. Also magnetization reversal can result from $\tau_L$, provided that the magnet respond on a time scale faster than the ferroelectric switching time scale; then magnetization is able to follow the rotation of the *L* vector (see Table 1).

We have fabricated Ferromagnet-Multiferroic heterostructure devices comprising of Current In-Plane Giant Magneto-Resistance (CIP-GMR) lateral spin valves ($Co_{90}Fe_{10}$/Cu/ $Co_{90}Fe_{10}$) coupled to multiferroic $BiFeO_3$ (Figure S1) (see Methods for device manufacturing processes). We deposit a multiferroic $(001)_p$ $BiFeO_3$ (100 nm) on a conducting layer $SrRuO_3$ (8 nm) on a (110) $DyScO_3$ substrate. $DyScO_3$ substrates provide an anisotropic strain in the $BiFeO_3$ film and allow only 2 of the 8 stable polarizations of the $BiFeO_3$ (Figure 3E, 3F show the PFM imaging of the 2-variant BFO). We use an electrode on the periphery of the chip to make contact to the $SrRuO_3$ conductive back-electrode for the application of out-of-plane electric field for polarization switching. A Giant Magneto-Resistance (GMR) stack is deposited *in-situ* on the $BiFeO_3$ under ultrahigh vacuum conditions to provide a direct sensing mechanism for the state of the electrically controlled bottom magnet. Three sets of test chips were processed comprising more than 1000 ME-GMR devices, which comprised GMR stripes at angles of 0, 45º, 90º and 135º with respect to the ferroelectric domains in the BFO (which lie along the pseudo-cubic $[100]_{pc}$ directions).

Using the sensitive and local CIP-GMR[27] magneto-transport measurements, we observe a magnetically stable exchange bias in devices with lateral dimensions ~200 nm (among the smallest devices made in our study). The exchange bias is manifested as an asymmetry in the resistance vs. applied magnetic field (R-H) hysteresis. The exchange bias is stable at room temperature with respect to magnetic cycling and progressively becomes stronger as the lateral dimensions are decreased to  below ~1-2 micrometers (Figure 3A). This observation is consistent with the local exchange bias observed in other systems viz.

Py/YMnO$_3$ [30] and Co/LaFeO3 [31], which show dimension dependent exchange bias [32]. In order to understand the fundamental origins of how the exchange bias emerges as the lateral dimensions are reduced, we carried out detailed micromagnetic simulations of the GMR stack, with various boundary conditions imposed on the stack. The micromagnetic (see Methods) effect of exchange coupling due to the antiferromagnetic order of BFO (**L**) and exchange bias due to canted magnetization (**M$_C$**) are shown in Figure S2, and compared with magnetic force microscopy images. All of the physical interaction parameters, such as the strength of exchange bias, the anisotropy, were kept unchanged, with the only variable being the lateral dimension of the ME-GMR element. The central results of our simulations of the GMR response as a function of magnetic field, for the two lateral length scales studied, are shown in Figs.3C, D. Consistent with the experimental data in Figs. 3A, B, the simulations also show an enhanced asymmetry of the 200nm wide GMR devices; this asymmetry progressively decreases as the lateral dimension is increased such that the 2000 nm devices show negligible shift. Factors relevant for switching are comprehended in the micromagnetic simulations: demagnetization, including dipole interaction between the top and bottom layers; material anisotropy; exchange stiffness; external magnetic field. Also coordinate dependent (due to striped domains) exchange bias (effective field) and exchange coupling (in-plane anisotropy) act on the bottom layer. Due to interface with Pt, the top layer is affected by out-of-plane anisotropy, which is however not sufficient to stabilize an out of plane magnetization state.

We performed micromagnetic simulations of the magneto-electric's exchange interaction with the GMR stack, to establish both the presence of a unidirectional anisotropy - exchange bias and the effect of electrical switching of the exchange bias. Micromagnetic simulations with locally modulated exchange interactions following the multiferroic domain pattern are performed for a) comparing ME-GMR device sizes (2000 nm X 200 nm Fig S3, 2000 nm X 1000 nm Fig S4) b) Reversal of the sign of exchange bias (Fig S5) c) Effect of domain orientation w.r.t to the ME-GMR device (45° orientation of ME-GMR w.r.t multiferroic domains (Fig S3), 135° orientation of ME-GMR w.r.t multiferroic domains (Fig S6)).

Micromagnetics show excellent agreement with the magneto-transport experimental data. The resulting net magnetization reversal of a 2000 nm X 200 nm spin valve layer in contact with the BFO is shown in Fig. S3, magnetization reversal of a 2000 nm X 1000 nm device is shown in S4. The magnetic patterns of the BFO are also visible in the top GMR layer due to exchange and dipole interactions within the CIP-GMR stack.  In the presence of the exchange bias, the hysteresis of magnetization vs. field (M-H) and, correspondingly, of magnetoresistance vs. field (R-H) change. The nature of this change is qualitatively different in wires of 2000x1000nm and 2000x200nm. For wires of 2000x200nm with the width smaller than the period of BFO domains, the shape anisotropy plays a significant role and contributes to a larger coercivity of the top layer. As magnetization switches with an external magnetic field, the magnetization in both layers remains approximately aligned along the FM wire, though with significant local deviations from it where forced by exchange coupling from BFO, see Figs. S3-S6. The switching to mostly the opposite direction happens as a sudden transition relative to the external field. We attribute this as the main reason for the sharp steps in the R-H hysteresis at well-separated values of applied field. The effect of the exchange bias coupling is that it would shift the M-H hysteresis loop of the bottom (but not the top layer) layer from the center position. At some value of the exchange bias, e.g. Heb=30Oe, these loops would overlap. However due to the dipole interaction between the top and bottom layer, they compete to switch first, see Figs. S3-S4. Depending on the parameters, and the direction of sweep, the top or the bottom layer switches first. This leads to large shifts of coercive fields and to asymmetric R-H hysteresis loops. In contrast, for the wires of 2000x1000nm, the shape anisotropy plays an insignificant role. Even in the presence of the exchange bias, the hysteresis loops for the top and bottom layers do not overlap. Also the magnetization rotates gradually as the external field is swept. All that leads to a smoother shape of the R-H hysteresis with little separation between branches.  Also the asymmetry of the hysteresis is hardly noticeable. In other words, we identify the large asymmetry of R-H hysteresis in wires narrower than the width of BFO domains (with lack of asymmetry in wider wires) as a signature of the exchange bias.

Armed with this critical insight that the lateral dimensions play a key role in the emergence of a measurable exchange bias[31], we then proceeded to ask: can this exchange bias be modulated with an electric field applied such that the BFO layer polarization is switched [32, 33]. The electrical control of the $BiFeO_3$ is obtained via contacts to a conductive oxide bottom electrode formed with $SrRuO_3$. The voltage controlled GMR hysteresis for the devices are shown in Figure 4. The GMR hysteresis of the electrically pristine devices (i.e., devices before the first ferroelectric switching was made is shown in figure 4A. A reversible exchange bias is observed with the application of an electric pulse (10 ms long, with a current limit of 10 uA). The two hysteresis loops indicate the magnetic switching of the bottom FM and top FM. The bottom FM has an enhanced magnetic anisotropy and switches at higher magnetic fields. In contrast, the top FM has lower anisotropy. We observe that the sign of the magneto-electric tuning of the hysteresis curves is consistent with a unidirectional anisotropy. Micro-magnetic simulations with exchange coupling following the multiferroic domain pattern is consistent with the reversal of exchange bias uniaxial anisotropy. In Fig S3, S5, consistent magneto-transport signature is observed, where the exchange bias reverses orientation when the FE polarization undergoes 180º reversal from [1,-1,1] to [-1,1,-1] (Figure S3, S5). However, the anti-ferromagnetic order (L) which is an axial quantity remains the same. We further compare the magneto-transport trends with the multiferroic micromagnetic simulations (Figures 4D,E,F) where a reversible unidirectional exchange anisotropy is applied to the free (bottom) layer (Figures S3, S5). To highlight the effect of the electric field on the exchange bias, we artificially centered the GMR hysteresis with an external field; for an external centering field of 200 Oe, for positive/negative voltages the switching field is enhanced/reduced. In contrast, when the centering field is reversed, the positive/negative voltages induce a reduction/enhancement of the switching fields. We note that this can be carried out reversibly for at least 100 switching cycles. This direct relationship between the polarity of the applied electric field and the exchange bias magnitude is perhaps the most important discovery in this work, since it directly demonstrates that the canted moment in the BFO is being switched by 180* by the application of an out of plane electric field .

We show magnetoelectric hysteresis loops of the multiferroic/ferromagnet heterostructure with voltage as the independent control signal and magnetization sensed with CIP-GMR as the output. The electrical test structure is shown in figure S1. The ME-GMR device is formed by a Ferromagnet-Multiferroic heterostructure devices comprising of Current In-Plane Giant Magneto-Resistance (CIP-GMR) lateral spin valves ($Co_{90}Fe_{10}$/Cu/ $Co_{90}Fe_{10}$) coupled to multi-ferroic $BiFeO_3$ (see Methods for device manufacturing processes). The devices are electrically addressable via a bottom contact made via SRO conducting back electrode. The magneto-electric transfer function with voltage is seen in figure 5A. Here we apply a switching voltage pulse to the bottom electrode of the ME-GMR test structure and measure the state of the magnet. We also measure the ferro-electric polarization as a function of the applied voltage (figure 5B). This ME-GMR hysteresis closely resembles the ferroelectric hysteresis. The magnitude of the electrically switched GMR hysteresis matches the GMR hysteresis via magnetic sweeps indicating a full 180º reversal of the FM layer in contact with BFO with respect to the top FM layer.

A direct consequence of a unidirectional anisotropy is the ability to project the exchange bias along a given vector, in contrast to uniaxial anisotropy. To test this hypothesis, we have fabricated devices with orientation at 45º and 135º to the striped domains on the BFO/SRO/DSO substrate (figures 5C, 5E). Here, the easy axis (length direction) of the ME-GMR device is aligned at 45º and 135º with respect to the BFO domain topology. We expect that a unidirectional exchange bias will project as a vector producing $B_{ex}$ proportional to cos(Θ) and cos(180º-Θ). Figure 5D, 5F demonstrate that the magneto-electric effect (exchange bias with respect to the applied electric field) indeed reverses direction, consistent with a uni-directional effect. We have further confirmed the effect of orientation of the with respect to the multiferroic domains via micro-magnetic simulations (compare Fig S3, S6). The asymmetry caused by the projection of the exchange bias along the easy axis of the device is reproduced in micro-magnetics, confirming that exchange bias due to interaction of the FM with the BFO indeed projects like a vector, with sign reversal when the projection direction changed (Fig. S6). Hence, we see a reversal in the sign of the magnetoelectricity depending on the orientation of the device with respect to the striped domains.

In conclusion, we observed a uni-directional exchange bias in a FM in contact with a room temperature multiferroic that is electrically switchable. An electrically controlled uni-directional exchange bias interacts with the magnet similar to a directional magnetic field and opens the possibility for a wider range of logic and memory devices. The direct relationship between the polarity of the applied electric field and the exchange bias is an important discovery, since it indicates that the uncompensated magnetization (canted moment in the BFO) is being switched by 180º by the application of an out of plane electric field. Demonstration of a local and directional magneto-electric switching mechanism can enable a wider class of logic/memory devices. We expect that the strength of the FM and Multi-ferroic interaction and magnetic switching to be improved with dimensional scaling and improved material quality. The ability to switch a nanomagnet with a locally reversible exchange field can enable a wide application space in memory and logic.

## METHODS

**Sample growth and process flow:** In this study, 20nm (001) SrRuO$_3$ / 100nm (001) BiFeO$_3$ were grown by pulsed-laser deposition using a KrF laser onto single-crystalline (110) DyScO$_3$ substrate. A striped ferroelectric domain structure with two polarization variants of 100nm BiFeO$_3$ film was confirmed by Piezoelectric Force Microscopy (PFM). A conducting 20nm SrRuO$_3$ acted a back electrode. Next, a GMR stack (2nm Co$_{0.9}$Fe$_{0.1}$/ 5nm Cu/ 2.5nm Co$_{0.9}$Fe$_{0.1}$ / 2nm Pt) was deposited by direct-current magnetron sputtering onto the 100nm BiFeO$_3$ film. A 2 nm Pt layer acted as a capping layer to prevent oxidation of other layers underneath. The GMR device was fabricated to study the voltage-controlled magneto-electric switching. The final device structure is shown in Fig S1.

**Micromagnetic Simulations:** Micromagnetic simulations are performed using the NIST's simulator OOMMF [27]. The inputs scripts for simulations are available among the supplementary files. The following parameters were chosen to model the device: $M_s = 1 MA/m$ - magnetization of CoFeB, $A_x = 10 pJ/m$ - exchange stiffness, $K_z = 200 kJ/m^3$ - uniaxial anisotropy in the top layer, $w_s = 200 nm$ - period of striped domains, $K_{ec} = 16 kJ/m^3$ - in-plane anisotropy in the bottom layer due to the exchange coupling, $t_b$ - thickness of the bottom layer. The patterns of exchange coupling axes and of exchange bias direction are shown in Fig. S2 (E). Magnetization is calculated by minimizing the overall energy of the system at every value of the external magnetic field. The magnetic field is swept from $H = 300 oe$ to $H = -300 oe$, and back.

**Magneto electric switching dynamics with exchange bias vs exchange coupling :** In exchange coupling mediated magneto-electric switching, the switching dynamics of the Ferromagnet (FM) have to closely follow the switching dynamics of the Multiferroic (MF). This is due to the uniaxial nature of the exchange coupling. In contrast, exchange bias mediated magneto-electric switching decouples the switching dynamics of the ferromagnet from the switching dynamics of the Multiferroic.

[1] Kuhn, Kelin J. "Considerations for ultimate CMOS scaling." *IEEE Transactions on Electron Devices* 59, no. 7 (2012): 1813-1828.

[2]  Ferain, Isabelle, Cynthia A. Colinge, and Jean-Pierre Colinge. "Multigate transistors as the future of classical metal-oxide-semiconductor field-effect transistors." Nature 479.7373 (2011): 310-316.

[3] Meindl, James D., Qiang Chen, and Jeffrey A. Davis. "Limits on silicon nanoelectronics for terascale integration." Science 293.5537 (2001): 2044-2049.

[4] Khomskii, Daniel. "Trend: Classifying multiferroics: Mechanisms and effects." *Physics* 2 (2009): 20.

[5] Spaldin, N. A. & Fiebig, M. The renaissance of magnetoelectric multiferroics. *Science* **309**, 391–392 (2005)

[6] Salahuddin, Sayeef, and Supriyo Datta. "Use of negative capacitance to provide voltage amplification for low power nanoscale devices." *Nano letters* 8, no. 2 (2008): 405-410.

[7] Manipatruni, Sasikanth, Dmitri E. Nikonov, and Ian A. Young. "Spin-orbit logic with magnetoelectric nodes: A scalable charge mediated nonvolatile spintronic logic." *arXiv preprint arXiv:1512.05428* (2015).

[8] Scott, J. F. Room-temperature multiferroic magnetoelectrics. *NPG Asia Mater.* **5**, e72 (2013)

[9] Wang, J. *et al.* Epitaxial BiFeO$_3$ multiferroic thin film heterostructures. *Science* **299**, 1719–1722 (2003)

[10] Chu, Y. H. *et al.* Low voltage performance of epitaxial BiFeO$_3$ films on Si substrates through La substitution. *Appl. Phys. Lett.* **92**, 102909 (2008)

[11] e.g Chung, et al. "A 3.3-V, 4-Mb nonvolatile ferroelectric RAM with selectively driven double-pulsed plate read/write-back scheme." *IEEE Journal of Solid-State Circuits* 35, no. 5 (2000): 697-704.

[12] M. Trentzsch *et al.*, "A 28nm HKMG super low power embedded NVM technology based on ferroelectric FETs," *2016 IEEE International Electron Devices Meeting (IEDM)*, San Francisco, CA, 2016, pp. 11.5.1-11.5.4.

[13] Ederer, Claude, and Nicola A. Spaldin. "Weak ferromagnetism and magnetoelectric coupling in bismuth ferrite." *Physical Review B* 71, no. 6 (2005): 060401.

[14] Heron, J. T., J. L. Bosse, Q. He, Y. Gao, M. Trassin, L. Ye, J. D. Clarkson et al. "Deterministic switching of ferromagnetism at room temperature using an electric field." *Nature* 516, no. 7531 (2014): 370-373.

[15] Allibe, J, et al, "Room temperature electrical manipulation of giant magnetoresistance in spin valves exchange-biased with BiFeO3." *Nano letters* 12, no. 3 (2012): 1141-1145.

[16] Wu, S. M., Shane A. Cybart, P. Yu, M. D. Rossell, J. X. Zhang, R. Ramesh, and R. C. Dynes. "Reversible electric control of exchange bias in a multiferroic field-effect device." *Nature materials* 9, no. 9 (2010): 756-761.

[17] Natarajan, S., M. Agostinelli, S. Akbar, M. Bost, A. Bowonder, V. Chikarmane, S. Chouksey et al. "A 14nm logic technology featuring 2 nd-generation FinFET, air-gapped interconnects, self-aligned double patterning and a 0.0588 μm 2 SRAM cell size." In Electron Devices Meeting (IEDM), 2014 IEEE International, pp. 3-7. IEEE, 2014.

[18] Nikonov, D.E.; Young, I.A., "Benchmarking of Beyond-CMOS Exploratory Devices for Logic Integrated Circuits," in Exploratory Solid-State Computational Devices and Circuits, IEEE Journal on , vol.1, no., pp.3-11, Dec. 2015

[19] Sun, Nian X., and Gopalan Srinivasan. "Voltage control of magnetism in multiferroic heterostructures and devices." In *Spin*, vol. 2, no. 03, p. 1240004. World Scientific Publishing Company, 2012.

[20] Landau, L. D., Lifshitz, E. M., and Pitaevskii, L. P. Electrodynamics of continuous media; Vol. 8 of the Course of Theoretical Physics. 2[nd] ed., Elsevier, 2013.[20] Wojdeł, Jacek C., and Jorge Íñiguez. "Magnetoelectric Response of Multiferroic BiFeO 3 and Related Materials from First-Principles Calculations." *Physical review letters* 103, no. 26 (2009): 267205.

[21] Stengel, Massimiliano, and Jorge Íñiguez. "Electrical phase diagram of bulk BiFeO 3." *Physical Review B* 92, no. 23 (2015): 235148.

[22] Vinokurov, D. L., and A. I. Morosov. "Magnetic structure of the compensated ferromagnet-multiferroic interface." *Physics of the Solid State* 55, no. 11 (2013): 2246-2251.

[23] Béa, H., M. Bibes, F. Ott, B. Dupé, X-H. Zhu, S. Petit, S. Fusil, C. Deranlot, K. Bouzehouane, and A. Barthélémy. "Mechanisms of exchange bias with multiferroic BiFeO 3 epitaxial thin films." *Physical review letters* 100, no. 1 (2008): 017204.

[24] Malozemoff, A. P. "Random-field model of exchange anisotropy at rough ferromagnetic-antiferromagnetic interfaces." *Physical review B* 35, no. 7 (1987): 3679.

[25] Wu, S. M., Shane A. Cybart, P. Yu, M. D. Rossell, J. X. Zhang, R. Ramesh, and R. C. Dynes. "Reversible electric control of exchange bias in a multiferroic field-effect device." *Nature materials* 9, no. 9 (2010): 756-761.

[26] He, Xi, et al. "Robust isothermal electric control of exchange bias at room temperature." *Nature materials* 9, no. 7 (2010): 579-585.

[27] M. J. Donahue and D. G. Porter, "OOMMF User's Guide, Version 1.0," National Institute of Standards and Technology Report No. NISTIR 6376, September 1999 (unpublished).

[28] 1J. C. Sankey, Y.-T. Cui, J. Sun, J. C. Slonczewski, R. A. Buhrman, and D. C. Ralph, Nat. Phys. 4, 67 (2008).

[29] H. Kubota, A. Fukushima, K. Yakushiji, T. Nagahama, S. Yuasa, K. Ando, H. Maehara, Y. Nagamine, K. Tsunekawa, D. D. Djayaprawira, N. Watanabe, and Y. Suzuki, Nat. Phys. 4, 37 (2008).

[30] Laukhin, Vladimir, Vassil Skumryev, X. Martí, D. Hrabovsky, F. Sánchez, M. V. García-Cuenca, C. Ferrater et al. "Electric-field control of exchange bias in multiferroic epitaxial heterostructures." Physical review letters 97, no. 22 (2006): 227201.

[31] Scholl, A., F. Nolting, Jin Won Seo, H. Ohldag, J. Stöhr, S. Raoux, J-P. Locquet, and J. Fompeyrine. "Domain-size-dependent exchange bias in Co∕ LaFeO 3." Applied physics letters 85, no. 18 (2004): 4085-4087.

[32] Martin, Lane W., Ying-Hao Chu, Qian Zhan, R. Ramesh, Shu-Jen Han, Shan X. Wang, Maitri Warusawithana, and Darrell G. Schlom. "Room temperature exchange bias and spin valves based on Bi Fe O 3∕ Sr Ru O 3∕ Sr Ti O 3∕ Si (001) heterostructures." Applied Physics Letters 91, no. 17 (2007): 172513.

[33] Allibe, J., Fusil, S., Bouzehouane, K., Daumont, C., Sando, D., Jacquet, E., Deranlot, C., Bibes, M. and Barthelemy, A., 2012. Room temperature electrical manipulation of giant magnetoresistance in spin valves exchange-biased with BiFeO3. Nano letters, 12(3), pp.1141-1145.

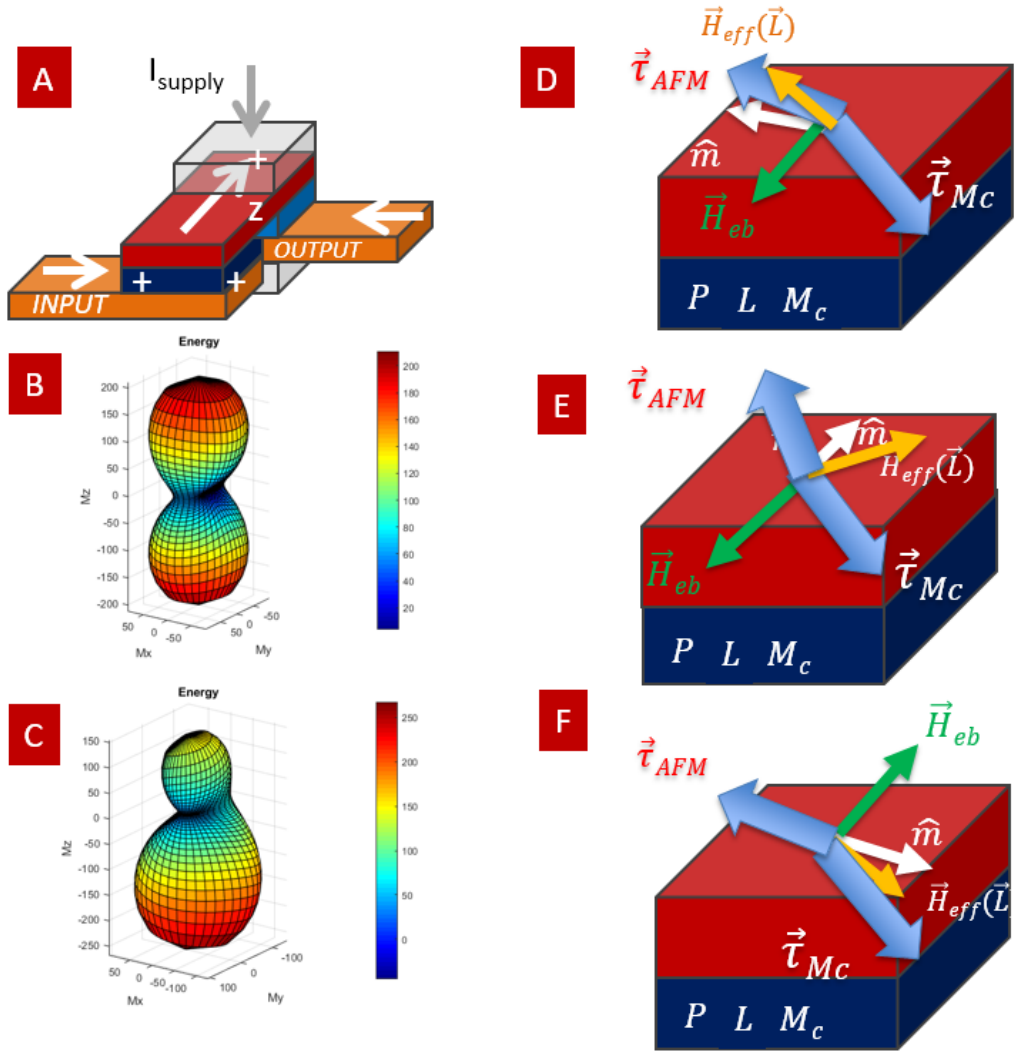

*Figure 1.* **Magento-electric torques on a ferromagnet** A) A magneto-electric spin logic device (MESO Logic [7]) showing the typical geometry where ME switching may be utilized to switch a nanomagnet for logic applications and ME voltage controleld magnetic memory. Magneto-static energy of the ferromagnet with B) Exchange Coupling only (symmetric about the inplane hard axis) C) Exchange bias and coupling (asymmetric about the hard axis). Exchange bias breaks the spatial symmetry in the energy landscape. Switching mechanism for a ferromagnet in contact with a Magneto-electric Multiferroic with multistep switching D) Initial State where $\tau_L$, TAFM & $\tau_{Mc}$, TMC and torque due to anisotropy produce a net zero ME torque. E) Intermediate state where the $\tau_L$, TAFM & $\tau_{Mc}$, TMC are driving the FM dynamics due to the intermediate polarization switching state of BiFeO$_3$ F) Final state where $\tau_L$, TAFM & $\tau_{Mc}$, TMC and torque due to anisotropy produce a net zero ME torque, $M_c=M_1+M_2$, $L=M_1-M_2$ experience 180º reversal.

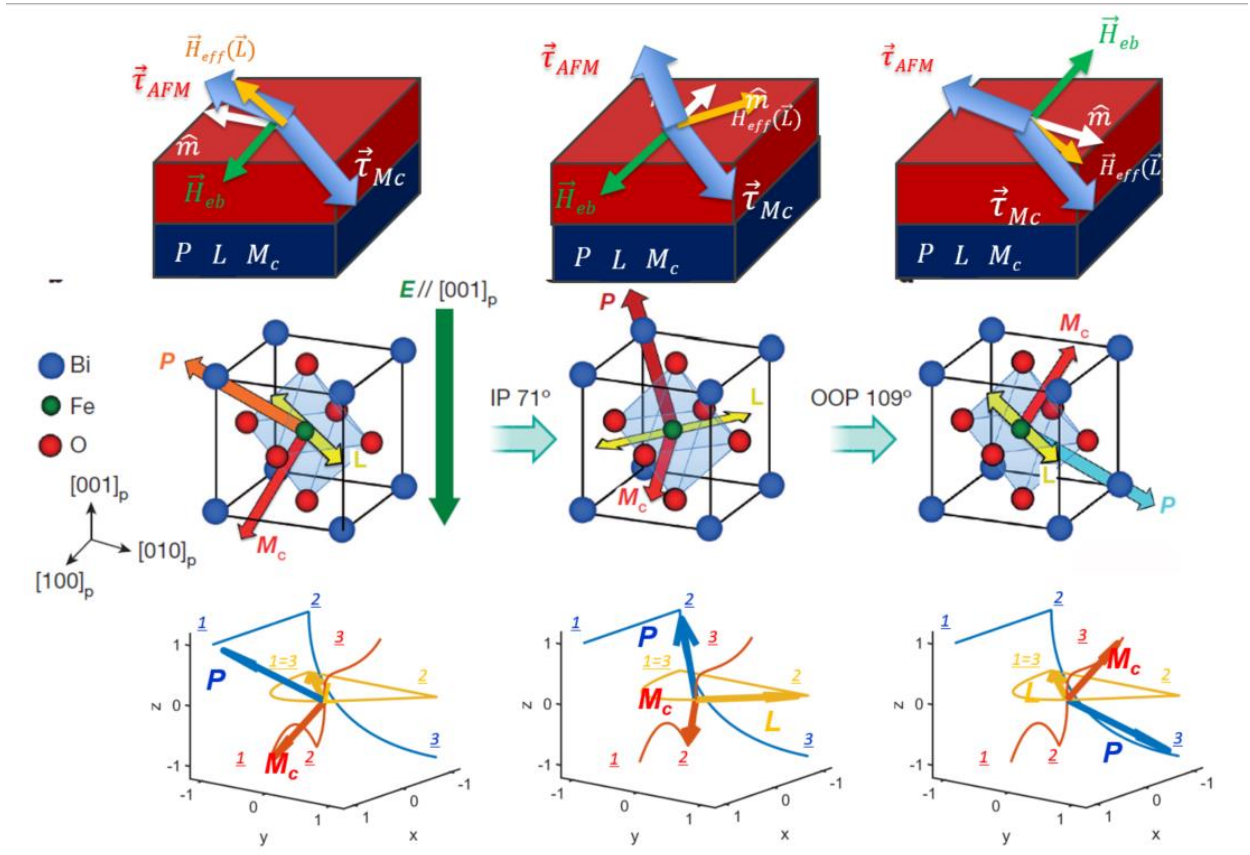

Figure 2. The geometry of magnetoelectric switching in the BFO/CoFe interface. Top row: the stack of BFO (blue) and CoFe (dark red) layers. Directions of magnetization ($\hat{m}$), effective magnetic field of exchange bias ($H_{eb}$) and of exchange coupling ($H_{eff}(L)$), torques due to exchange bias ($\tau_{Mc}$) and exchange coupling ($\tau_{AFM}$). Middle row: directions of polarization ($P$), antiferromagnetic order ($L$), and canted magnetization ($M_c$) relative to ions in a crystal lattice cell of BFO. Bottom row: trajectories of $P$, $L$, and $M_c$ relative to the crystal axes. In columns: three special states in the switching process: the initial, after a 71° turn of polarization, final after a 109° turn.

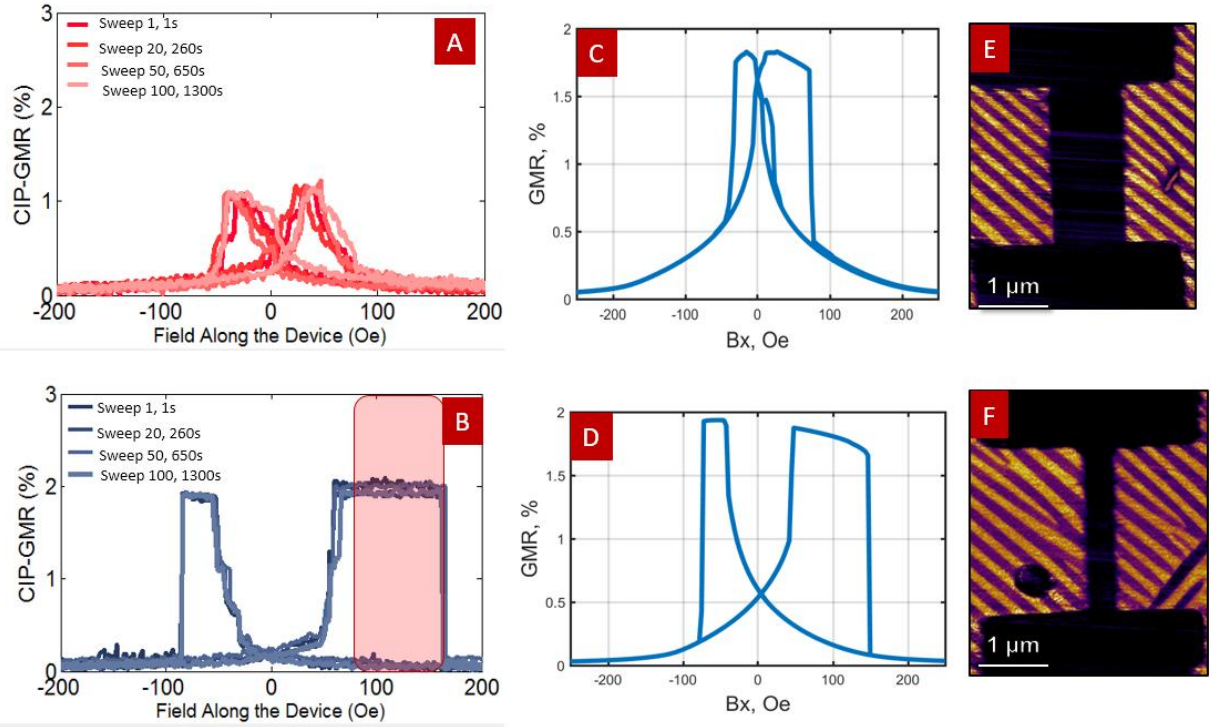

*Figure 3. **Exchange bias in Magneto-electric GMR devices**. Magnetic-transport measurement of an ME-GMR device (geometry and cross section are shown in Fig S2) fabricated on a 2-variant multiferroic BFO with A) lateral dimensions 1 μm X 2 μm B) lateral dimensions 200 nm X 2 μm. Micromagnetic simualtion of the ME-GMR device comprehending the exchange bias and exchange coupling from multiferroic substrates C) for lateral dimensions 1 μm X 2 μm D) 200 nm X 2 μ. PFM images of the devices for dimensions E) 1 μm X 2 μm F) 200 nm X 2 μm. Scaled devices show presence of an exchange bias exerted on the FM layer in contact with BFO. Figures A,B show the Magnetic transport measurement repared for 100 cycles (1300 s).*

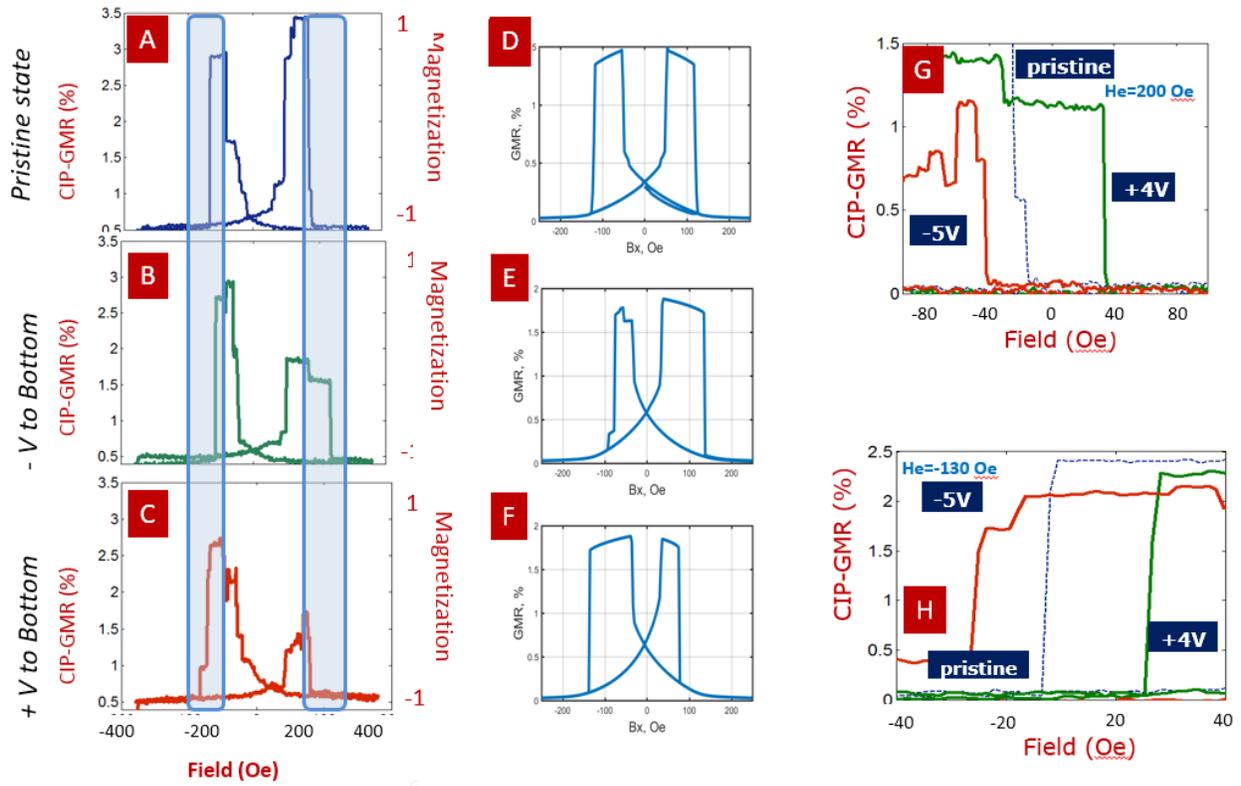

Figure 4. **Reversible Exchange bias in Magneto-electric GMR devices**: A) Prestine GMR devices before first switching of ferro-electricity showing no asymmetry in the GMR transport B) ME-GMR with negative exchange bias, which produces a higher positive switching field C) ME-GMR with postive exchange bias, which produces a higher negative switching field for the contact ferromagnet; Micromagnetic simualtion of ME-GMR devices D) without exchange bias before first FE switching of the device E) with negative exchange bias (see Fig S5) F) with positive exchange bias (see Fig S4) G) ME-GMR measurement with an external applied field of 200 Oe H) ME-GMR measurement with the an external applied field of -130 Oe. Shift in switching field of the FM in contact with BFO is consitent with voltage control of the exchange bias.

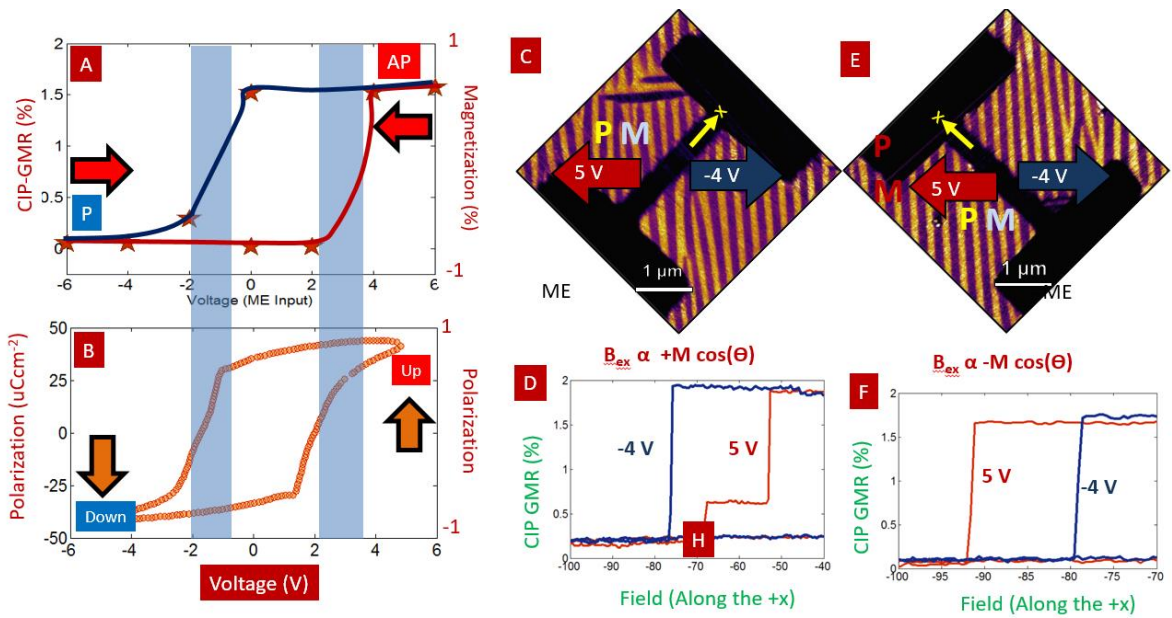

Figure 5. **Switching of magnetization by voltage.** A) Magneto-electric controlled GMR hysteresis showing ~full GMR switching of the device B) Ferro-electric hysteresis loop of the test structure. We note that the ME-GMR and PE loops have similar switching voltages C) ME-GMR devices fabricated at an orientation of +45º w.r.t to striped domains of BFO D) Magneto-electric tuning of the ME-GMR stucture shown in C. E) ME-GMR devices fabricated at an orientation of 135º w.r.t to striped domains of BFO F) Magneto-electric tuning of the ME-GMR structure shown in E.

| Magnetoelectric torques | Antiferromagnetic Torque | Exchange Bias Torque |
|---|---|---|
| Origin | Coupling to L of BFO | Coupling to $M_c$ of BFO |
| Directionality/Symmetry | Uniaxial Anisotropy ($L^2$) | Unidirectional Anisotropy ($M_c$) |
| Effective field | In-plane effective field (parallel to the L) | Canted, with an out-of-plane component |
| Steady state effect | Coercivity enhancement | Hysteresis shift |
| Magnetization reversal | Kinetic process (multi-step) where it follows the L vector | Independent switching trajectory, determined by $M_c$ |

Table 1. Comparisons of the two factors in the magnetoelectric switching, antiferromagnetic exchange coupling, and exchange bias.

# Supplementary Materials to

# "Electrical Control of Exchange Bias in Ferromagnet-Multiferroic Devices at Room Temperature"

S. Manipatruni[1], D. Nikonov[1], C.C. Lin[1], L. Dedon[2], A.R. Damodaran[2], Z. Chen[2], P. Bhagwati[2], Y. Lin[2], R. Ramesh[2] and I. Young[1]

[1]Components Research, Intel Corp., Hillsboro, Oregon, 97124, USA

[2]Department of Materials Science and Engineering, University of California, Berkeley, California 94720, USA

## Section S1. Physics of exchange bias

BFO, a multiferroic magnetoelectric is characterized by polarization **P**, antiferromagnetic order **L**, and canted magnetization **M_c**; all mutually perpendicular [i]. Polarization of BFO (**P**) is oriented to the vertices of the cube and antiferromagnetic order (**L**) points at the sides of the cube, the canted magnetization (**M_c**) points along the vector product $M_c \propto [P \times L]$ (see Figure 2). Domain walls in BFO are defined by the change of polarization between their two sides. In these experiments, we encounter **P** changing by 71°. The projection of **P** perpendicular to the wall remains the same, the projection of **P** parallel to the wall switches to opposite. Canted magnetization (**M_c**) changes between the two sides of a domain wall in a similar manner. The spatial period in the pattern of magnetization (comprising two stripes) has a typical value of 300 nm as observed in PFM imaging.

Magnetoelectric effect encountered in our experiments is actuated by electric field E perpendicular to plane. When the voltage is reversed, all BFO order parameters – **P**, **L**, **M_c** switch to opposite. Reversal of **P**, **L** and **M_c** occurs in every striped domain [ii]. The exchange interactions at the interface of BFO and the ferromagnet affect its evolution. While **L** contributes to an increase of in-plane anisotropy ("exchange coupling"), **Mc** contributes to exchange bias (i.e. local effective magnetic field). The differences in the effects of L and Mc are summarized in table 1.

Overall, in the two-step process, the direction of polarization **P** is reversed, and so is the direction of canted magnetization defining the exchange bias. An ambiguity exists about the direction of the antiferromagnetic order due to the opposite spin directions at the two sublattices. Thus the distribution of these spins would be equivalent if the antiferromagnetic order is switched by either 180° or by 360°. Using the relation [iii] between the polarization $P = (P_{inplane}; P_{outplane})$ and the antiferromagnetic order $L \propto P_{inplane} \times P_{outplane}$, we obtain the trajectory of these orders in the switching process, see Figure 2. The result of this consideration is that the antiferromagnetic order is switched in a closed trajectory by 360°. The character of this trajectory is such that, in the first step and part of the second step, the antiferromagnetic order $L$ turns by 180° while maintaining a large absolute value. In the remaining part of the trajectory, it performs another 180° turn while decreasing in the absolute value, approaching zero, and then increasing in value.

## Section S2. Magnetic hysteresis of the stack of layers

Factors relevant for switching are comprehended in the micromagnetic simulation: demagnetization, including dipole interaction between the top and bottom layers; material anisotropy; exchange stiffness; external magnetic field. Also coordinate dependent (due to striped domains) exchange bias (effective field) and exchange coupling (in-plane anisotropy) act on the bottom layer. Due to interface with Pt, the top layer is affected by out-of-plane anisotropy, which is however not sufficient to stabilize magnetization out of plane, see Figure S2.

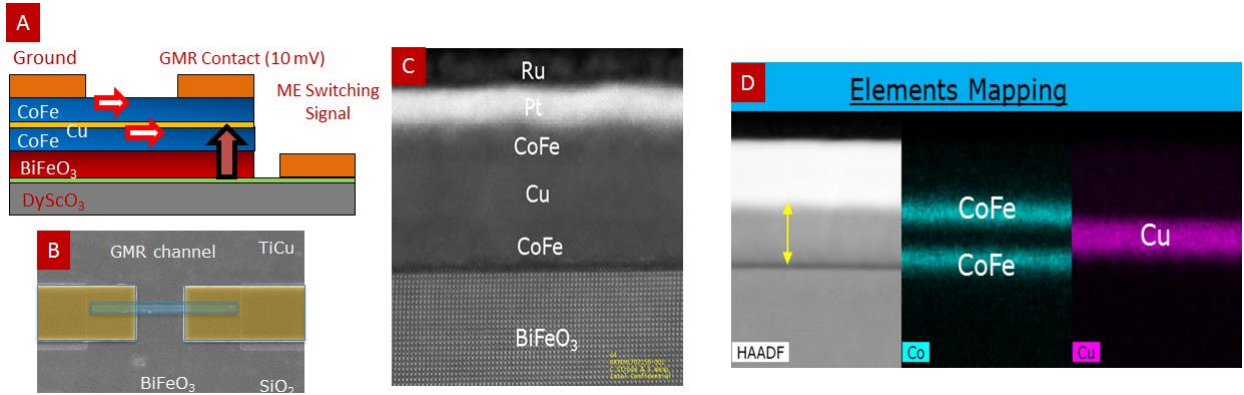

Figure S1. . The patterned device geometry. (A) Cross-section of the devices showing the DSO substrate, the BFO/CoFeB interface and the GMR stack (CoFe/Cu/CoFe) as well as electrodes. (B) Top view micrograph showing the GMR conducting channel. (C) Cross-section TEM micrograph of BFO and the GMR stack. (D) HAADF cross-section image and an overlap with the SIMS element maps.

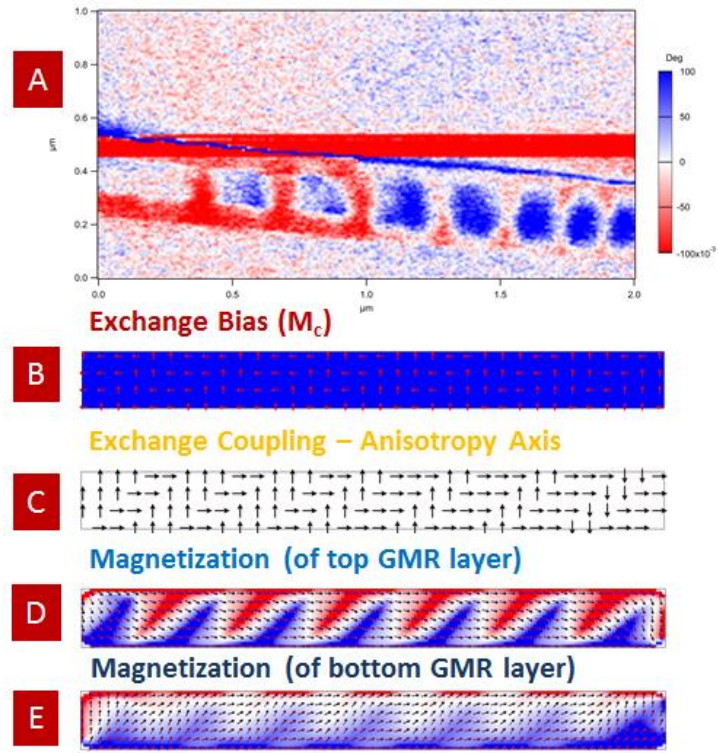

Figure S2. (A). MFM image of a GMR stripe. Color corresponds to the vertical projection of magnetization in the top layer, blue = up, red = down. (B) Map of the exchange bias magnetic field (proportional to $M_c$) in the bottom ferromagnetic layer. (C) Magnetic anisotropy (proportional to $L$) in the bottom ferromagnetic layer. (D) Simulated magnetization distribution in the top ferromagnetic layer. Arrows = in-plane projections of magnetization. Color =

out-of-plane projection of magnetization, blue = down, red = up. (E) Distribution of magnetization in the bottom ferromagnetic layer. The width of domains is $w_s$=300nm.

The magnetic field is first swept to positive, then negative, and back to positive values along the axis of the GMR conductor (see Figure S3). Magnetization is switched in this process in both top and bottom layers. Very high field aligns magnetization to its direction. At a smaller field, magnetization in the bottom layer tends to point along the local exchange coupling axis in the corresponding stripe, this direction is affected by exchange bias. Thereby BFO domains imprint on the ferromagnetic domains. Magnetization shows an oscillatory pattern corresponding to the stripes of the domains.

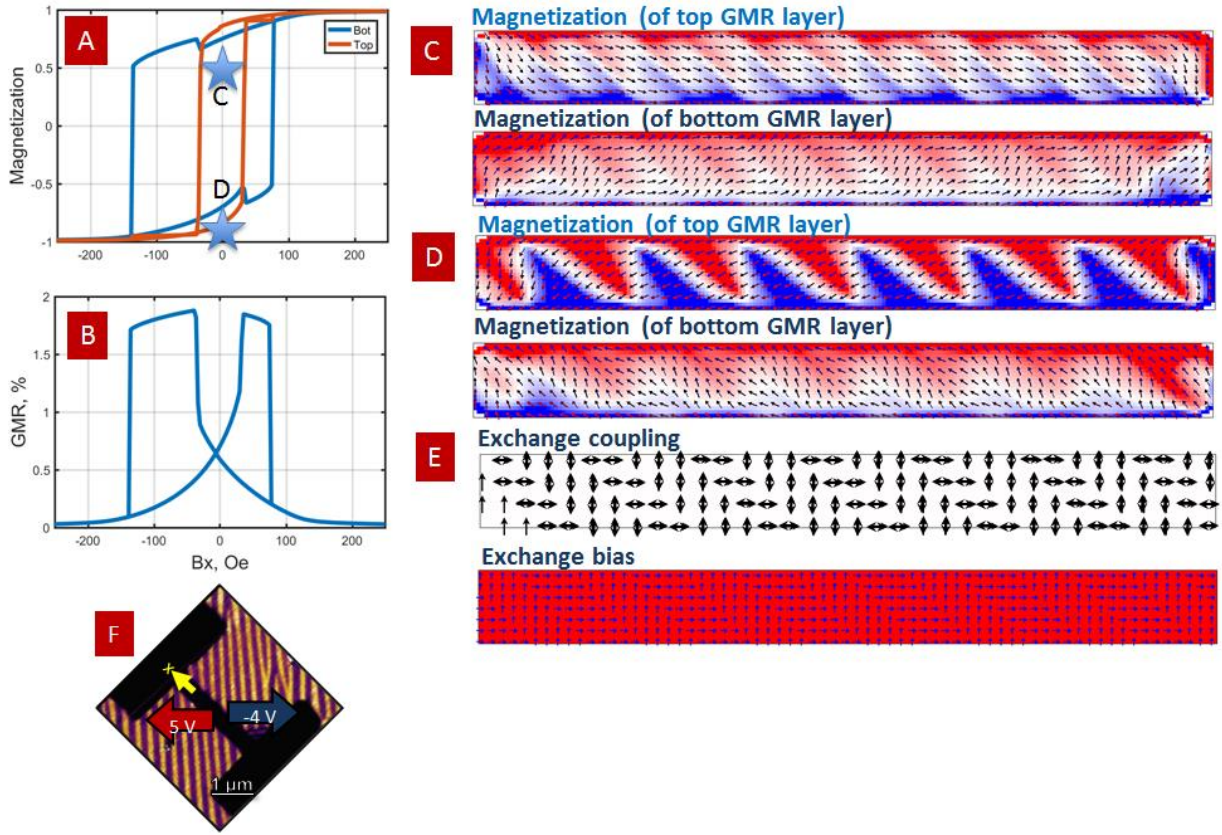

Figure S3. Magnetization Patterns in the Hysteresis Loop, for a devices of 2000x200nm area, applied voltage 5V, exchange bias $H_{eb} = 30 oe$. Domain stripes at -45° to the long axis. (A) Magnetizations (units of Ms) in the top and bottom layers of the spin valve vs. external field. (B) Giant magnetoresistance hysteresis in the spin valve. (C) Patterns of magnetizations in the top and bottom layers at external field H=0 on the sweep down. (D) Patterns of magnetizations in the top and bottom layers at external field H=0 on the sweep up. (E) Map of exchange coupling axes and exchange bias directions in the bottom layer. Arrows designate projections on the plane of the device. Color designates out-of-plane projections: red=positive, blue=negative. Color scale: the boundary of bright red corresponds to Mz=10kA/m. (F) PFM image of the ferroelectric domains. Black areas correspond to the GMR conductor and the electrodes. Directions of the applied voltage shown.

Out-of-pane (z) projection of magnetization exists especially at the edges, but also in domains, see Figure S2. Due to dipole interaction, domains from the bottom layer imprint on the top layer. The z-projection of magnetization in the top layer is larger due to out-of-plane anisotropy partially compensating demagnetization and due to lack of in-plane anisotropy. It is still in the order of 0.01 of the total Ms. Due to these factors of anisotropy, the coercive field is in general smaller for the top layer than for the bottom layer. In the presence of the exchange bias, the hysteresis of magnetization vs. field (M-H) and, correspondingly, of magnetoresistance vs. field (R-H) change. The character of this change is qualitatively different in wires of 2000x1000nm and 2000x200nm.

Using micromagnetic simulations comprehending the vector and local nature of the exchange coupling from the multi-ferroic, we identify that the large asymmetry of R-H hysteresis in ME-GMR devices narrower than the width of BFO domains as a signature of the exchange bias. We performed micromagnetic simulations of 2000 nm X 200 nm ME-GMR devices, where the width of the device is smaller than the period of the BFO domains.

The interplay of the vector contributions of exchange bias, exchange coupling and shape anisotropy lead to clear magneto-electric transport measurement indicating the presence of exchange bias. For wires of 2000x200nm with the width smaller than the period of BFO domains, the shape anisotropy plays a significant role and contributes to a larger coercivity of the top layer. As magnetization switches in the sweep of external magnetic field, the magnetization in both layers remains approximately aligned along the FM wire, though with significant local deviations from it where forced by exchange coupling from BFO, see Figures S3. The switching to mostly the opposite direction happens as a sudden transition relative to the external field. That is the reason for sharp steps in the R-H hysteresis at well separated values of the field. The effect of the exchange bias is that it would shift the M-H hysteresis loop of the bottom (but not the top layer) layer from the center position. This leads to a large shifts of coercive fields and to a significant asymmetry of the R-H hysteresis loops.

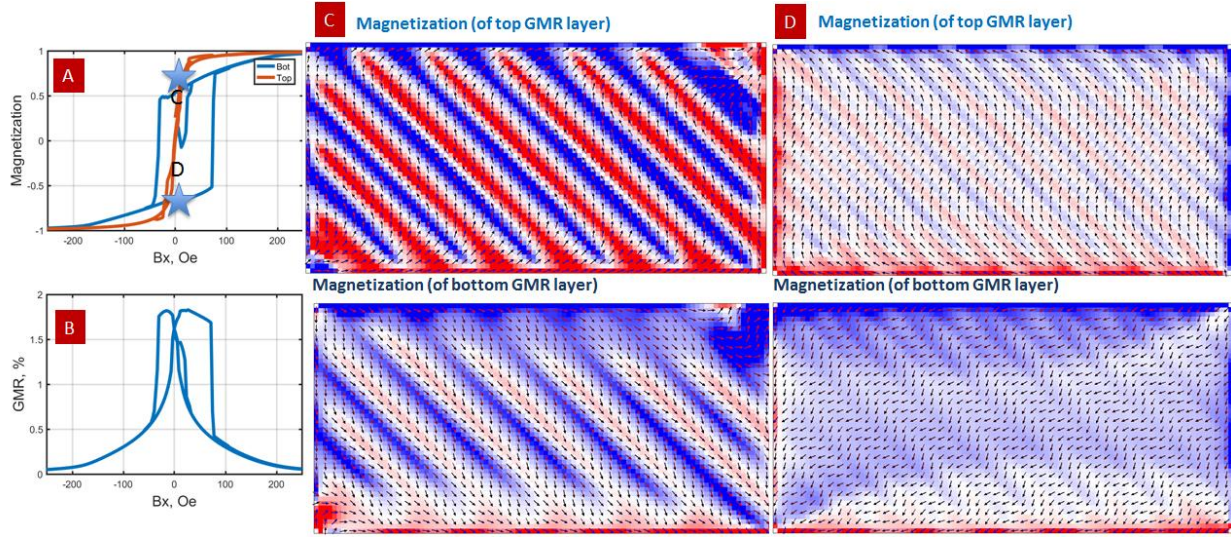

Figure S4. Effect of Ferromagnet Width, for a devices of 2000x1000nm area, exchange bias $H_{eb} = 30oe$. Domain stripes at -45° to the long axis. Domain stripes at -45° to the long axis. (A) Magnetizations (units of Ms) in the top and bottom layers of the spin valve vs. external field. (B) Giant magnetoresistance hysteresis in the spin valve. (C) Patterns of magnetizations in the top and bottom layers at external field H=0 on the sweep down. (D) Patterns of magnetizations in the top and bottom layers at external field H=0 on the sweep up.

In contrast, for the wires of 2000x1000nm, the shape anisotropy plays an insignificant role. Therefore the coercivity of the top layer is quite small, see Figure S4. Even in the presence of the exchange bias, the hysteresis loops for the top and bottom layers do not overlap. Also the magnetization turns gradually as the external field is swept. All that leads to a smoother shape of the R-H hysteresis with little separation between branches. Also the asymmetry of the hysteresis is hardly noticeable. In other words, we identify the large asymmetry of R-H hysteresis in ME-GMR devices narrower than the width of BFO domains (with lack of asymmetry in wider wires) as a signature of the exchange bias. Note that the magnetization in layers is more uniform and turns gradually, e.g. to +/-90° before it switched to the opposite direction.

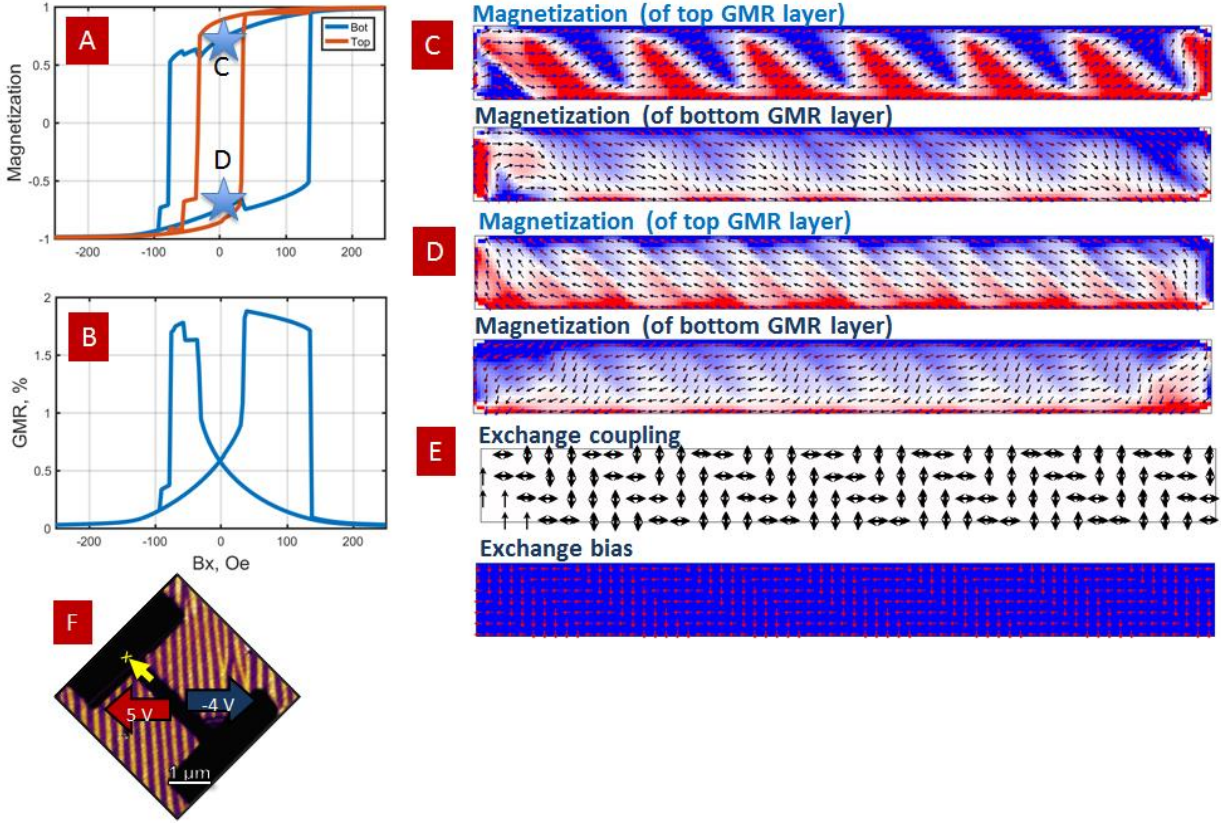

Figure S5. Switching of Exchange Bias Sign, for a devices of 2000x200nm area, applied voltage -4V, exchange bias $H_{eb} = -30 oe$. Domain stripes at -45° to the long axis. (A) Magnetizations (units of Ms) in the top and bottom layers of the spin valve vs. external field. (B) Giant magnetoresistance hysteresis in the spin valve. (C) Patterns of magnetizations in the top and bottom layers at external field H=0 on the sweep down. (D) Patterns of magnetizations in the top and bottom layers at external field H=0 on the sweep up. (E) Map of exchange coupling axes and exchange bias directions in the bottom layer. Arrows designate projections on the plane of the device. Color designates out-of-plane projections: red=positive, blue=negative. Color scale: the boundary of bright red corresponds to Mz=10kA/m. (F) PFM image of the ferroelectric domains. Black areas correspond to the GMR conductor and the electrodes. Directions of the applied voltage shown.

Next we simulated the impact of reversal of exchange bias that the multi-ferroic exerts on the FM. In Figure S5, we have reversed the direction of the exchange bias (compare Figure S5E with S3E). The measured magneto-electric transport data (Fig 4) is consistent with the model that the exchange bias from the underlying BFO is reversed. The hysteresis loops of the bottom FM, directly in contact with the BFO are approximately mirror symmetric relative to the vertical axis compared to Figure S3. Similarly, the patterns of magnetization are symmetric relative to the reversal of all coordinates. Hence, the micro-magnetics comprehending the vector interaction of the multi-ferroic's exchange coupling to FM are consistent with the electrical tuning of the exchange bias observed in figure 4.

Next, we simulated the impact of the orientation of the device with respect to the multi-ferroic domain structures. Uni-directional anisotropy is expected to produce a reversal in the sign of magneto-electric effect since it projects as a vector. The case of a ferromagnetic wire on the BFO layer as in Figure S3 but turned by 90° is shown in Figure S6. On the average the exchange bias points at -45° relative to the long x-axis of the FM wire. The R-H hysteresis is mirror symmetric to Figure S3. The patterns of magnetization are mirror symmetric relative to the X-axis.

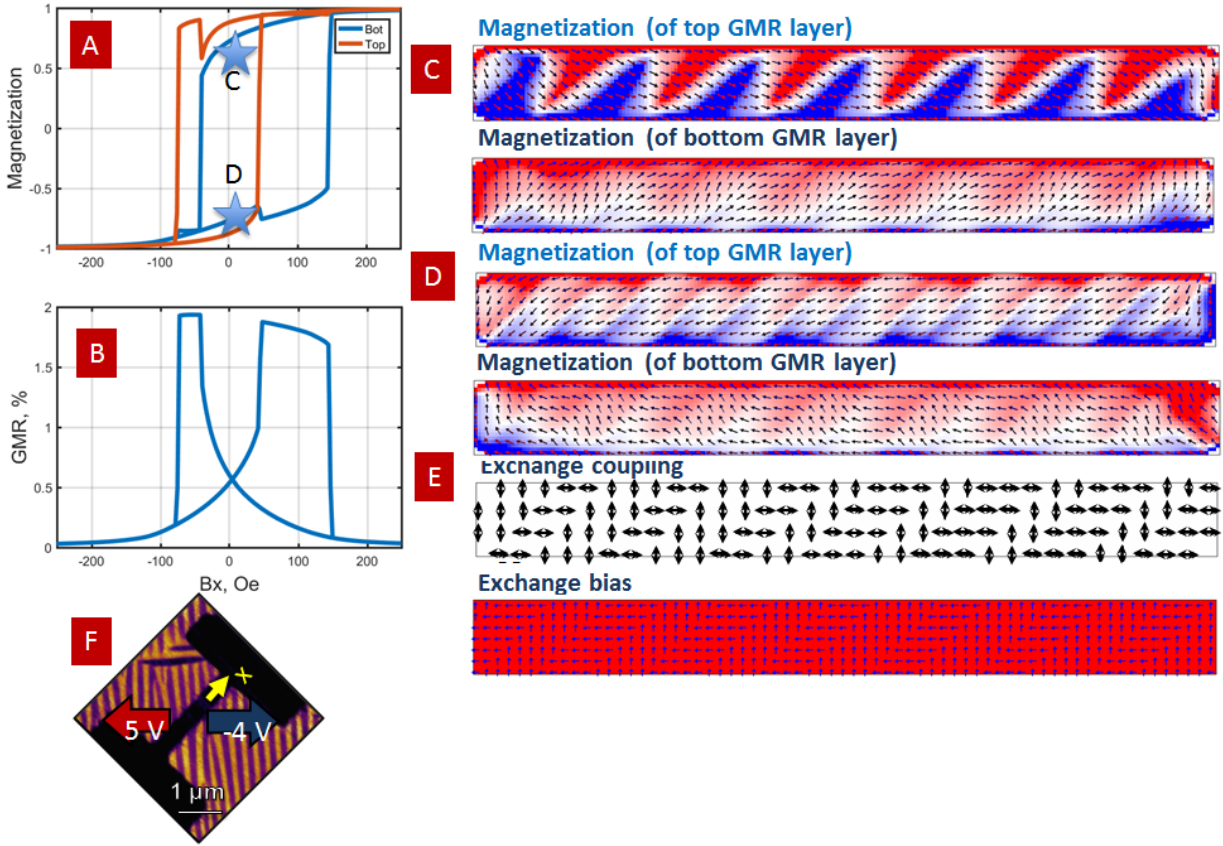

Figure S6. Effect of Domain Orientation in BFO, for a devices of 2000x200nm area, applied voltage 5V, exchange bias $H_{eb} = -30 oe$ (due to a different orientation of the electric field and the magnetic field (x=axis). Domain stripes at +45° to the long axis. (A) Magnetizations (units of Ms) in the top and bottom layers of the spin valve vs. external field. (B) Giant magnetoresistance hysteresis in the spin valve. (C) Patterns of magnetizations in the top and bottom layers at external field H=0 on the sweep down. (D) Patterns of magnetizations in the top and bottom layers at external field H=0 on the sweep up. (E) Map of exchange coupling axes and exchange bias directions in the bottom layer. Arrows designate projections on the plane of the device. Color designates out-of-plane projections: red=positive, blue=negative. Color scale: the boundary of bright red corresponds to Mz=10kA/m. (F) PFM image of the ferroelectric domains. Black areas correspond to the GMR conductor and the electrodes. Directions of the applied voltage shown.

## Section S3. Input script for the OOMMF simulator.

# MIF 2.1
# hysteresis of a ferromagnet under exchange bias from BFO
# in-plane magnetization
# rectangular nanomagnet
# stripes of exchange bias
# constants

```
set pi [expr 4*atan(1.0)]                              ;# number pi
set mu0 [expr 4*$pi*1e-7]                              ;# premeability of vacuum, kg*m/C^2
Parameter Temp 300                                     ;# temperature in K
set seed [expr int(714025.0 * rand())]
# material parameters
Parameter Ms 1000e3                                    ;# sat magnetization, A/m
Parameter Ms_r 1200e3                                  ;# sat magnetization, A/m
Parameter Ax  1e-11                                    ;# exchange constant, J/m
Parameter alp 0.01                                     ;# Gilbert damping
Parameter Lambda 1.0                                   ;# angle dependence of spin torque
Parameter K1 0e4                                       ;# cubic anisotropy, J/m^3
Parameter Ku_b 0e5                                     ;# one axis anisotropy, J/m^3
Parameter Ku_t 2e5                                     ;# one axis anisotropy, J/m^3
Parameter Heb0 30                                      ;# Exchange bias field, Oe
set Heb [expr {$Heb0*1e-4/$mu0}]                       ;# Convert to A/m
set ws 300e-9                                          ;# width of two stripes, m
set vec1 [list 1.0 0.0 1.4]                            ;# vectors of exchange bias
set vec2 [list 0.0 1.0 1.4]
set cec1 [list 0.0 1.0 0.0]                            ;# vectors of exchange coupling
set cec2 [list 1.0 0.0 0.0]
Parameter Kec 1.6e4                                    ;# exchange coupling energy w BFO, J/m^3
# geometry
Parameter asp 10                                       ;# aspect ratio
Parameter Lnm 200                                      ;# feature size, nm
set L [expr 1e-9*$Lnm]                                 ;# feature size, m
set width  [expr $L]
set length [expr {$asp*$width}]
set thick   2e-9
Parameter xycellsize 20e-9
set zcellsize [expr {1*$thick}]
set z1 [expr {1*$thick}]
set z2 [expr {$thick+$z1}]
set z3 [expr {$thick+$z2}]
Parameter curad 30e-9                                  ;# radius of rounded corners, m
Parameter asprad 2                                     ;# aspect ratio of the rounded corner
# device parameters
Parameter Happ 300.0                                   ;# External field, Oe
set Happ [expr {$Happ*1e-4/$mu0}]                      ;# Convert to A/m
Parameter Hdip 0.0                                     ;# Dipole field, Oe
set Hdip [expr {$Hdip*1e-4/$mu0}]                      ;# Convert to A/m
Parameter co 1.0                                       ;# cosine along main direction
Parameter devx 0.0                                     ;# deviation in x
Parameter devy 0.0                                     ;# deviation in y
Parameter devz 0.0                                     ;# deviation in z
# vector of initial magnetization
Parameter mo_theta 0.0                                 ;# Direction of mo, in degrees
set mo_theta [expr {$mo_theta*$pi/180.}]
Parameter mo_phi 0.0                                   ;# Direction of mo, in degrees
set mo_phi [expr {$mo_phi*$pi/180.}]
set ovect [list [expr {cos($mo_theta)}] [expr {sin($mo_theta)*cos($mo_phi)}] [expr {sin($mo_theta)*sin($mo_phi)}]
]
# execution options
set basename [subst hy$Lnm ]
```

```
Specify Oxs_BoxAtlas:freelay [subst {
  xrange {0 $length}
  yrange {0 $width}
  zrange {0 $z1}
}]
Specify Oxs_BoxAtlas:spacer [subst {
  xrange {0 $length}
  yrange {0 $width}
  zrange {$z1 $z2}
}]
Specify Oxs_BoxAtlas:topref [subst {
  xrange {0 $length}
  yrange {0 $width}
  zrange {$z2 $z3}
}]
Specify Oxs_MultiAtlas:vsyo {
atlas :freelay
atlas :spacer
atlas :topref
}
Specify Oxs_MultiAtlas:magnets {
atlas :freelay
atlas :topref
}
Specify Oxs_RectangularMesh:mesh [subst {
  cellsize {$xycellsize $xycellsize $zcellsize}
  atlas :vsyo
}]
# Geometry of the output electrode
Specify Oxs_ScriptScalarField:OutputEl [subst {
        atlas :vsyo
        script {Poln 1}
        script_args relpt
}]
Specify Oxs_AtlasScalarField:OutputEl_f [subst {
        atlas :vsyo
        default_value 0
        values {
                freelay :OutputEl
        }
}]
Specify Oxs_AtlasScalarField:OutputEl_r [subst {
        atlas :vsyo
        default_value 0
        values {
                topref :OutputEl
        }
}]
Specify Oxs_AtlasScalarField:Ktot [subst {
   atlas :magnets
   default_value 0
   values {
      freelay $K1
```

```
                    topref $K1
            }
}]
# Cubic anisotropy
Specify Oxs_CubicAnisotropy [subst {
  K1  :Ktot
  axis1 {1 0 0}
  axis2 {0 1 0}
}]
Specify Oxs_AtlasScalarField:Kunitot [subst {
    atlas :magnets
    default_value 0
    values {
       freelay $Ku_b
                    topref $Ku_t
          }
}]
# Uniaxial anistropy.
Specify Oxs_UniaxialAnisotropy:pma [subst {
   K1 :Kunitot
   axis {0 0 1}
}]
Specify Oxs_Exchange6Ngbr [subst {
default_A 0.0
atlas magnets
A {
freelay freelay $Ax
topref topref $Ax
}
}]
# Demag
Specify Oxs_Demag {}
Specify Oxs_ScriptVectorField:stripeEB [subst {
         script {StripesLim $ws $vec1 $vec2}
         script_args {relpt span}
         atlas :freelay
}]
# exchange bias field
Specify Oxs_FixedZeeman:exchbias [subst {
    field :stripeEB
    multiplier $Heb
}]
Specify Oxs_ScriptVectorField:stripeEC [subst {
         script {Stripes $ws $cec1 $cec2}
         script_args {relpt span}
         atlas :freelay
}]
Specify Oxs_AtlasScalarField:Kectot [subst {
    atlas :magnets
    default_value 0
    values {
       freelay $Kec
          }
```

```
}]
Specify Oxs_UniaxialAnisotropy:coeenh [subst {
   K1 :Kectot
   axis :stripeEC
}]
# field with a slight angle to x
Specify Oxs_UZeeman [subst {
  Hrange {
{ 0 0 0 [expr {$Happ*$co}] [expr {$Happ*$devy} ] [expr {$Happ*$devz}]   100 }
{ [expr {$Happ*$co}] [expr {$Happ*$devy}] [expr {$Happ*$devz}] [expr {-$Happ*$co}] [expr {-$Happ*$devy}] [expr {-$Happ*$devz}]   200 }
{ [expr {-$Happ*$co}] [expr {-$Happ*$devy}] [expr {-$Happ*$devz}] [expr {$Happ*$co}] [expr {$Happ*$devy} ] [expr {$Happ*$devz}]   200 }
  }
}]
# Magnetization
Specify Oxs_ScriptScalarField:Ms [subst {
        script {RoundedCorners $Ms $curad $asprad}
        script_args {relpt span}
        atlas :freelay
}]
Specify Oxs_ScriptScalarField:Msr [subst {
        script {RoundedCorners $Ms_r $curad $asprad}
        script_args {relpt span}
        atlas :topref
}]
Specify Oxs_AtlasScalarField:Mstot [subst {
   atlas :vsyo
   default_value 0
   values {
      freelay :Ms
                topref :Msr
        }
}]
Specify Oxs_AtlasScalarField:alp [subst {
   atlas :vsyo
   default_value 1
   values {
      freelay $alp
                topref $alp
        }
}]
# initial state
Parameter Input ""
if { [string length $Input] > 0 } {
   # we'll assume that readability has been checked externally
   Specify Oxs_FileVectorField:init [ subst {
        file $Input
        atlas :magnets
   } ]
} else {
        Specify Oxs_AtlasVectorField:init [ subst {
        atlas :magnets
```

```
                norm 1
                default_value {0. 0. 0.}
                values {
                        freelay {[lindex $ovect 0] [lindex $ovect 1] [lindex $ovect 2]}
                        topref {[lindex $ovect 0] [lindex $ovect 1] [lindex $ovect 2]}
                        }
                } ]
}
# projection fields for output
Specify Oxs_MaskVectorField:mxout_f {
        field {1 0 0}
        mask :OutputEl_f
}
Specify Oxs_MaskVectorField:myout_f {
        field {0 1 0}
        mask :OutputEl_f
}
Specify Oxs_MaskVectorField:mzout_f {
        field {0 0 1}
        mask :OutputEl_f
}
Specify Oxs_MaskVectorField:mxout_r {
        field {1 0 0}
        mask :OutputEl_r
}
Specify Oxs_MaskVectorField:myout_r {
        field {0 1 0}
        mask :OutputEl_r
}
Specify Oxs_MaskVectorField:mzout_r {
        field {0 0 1}
        mask :OutputEl_r
}
# start the simulation
   # Evolver
   Specify Oxs_CGEvolve:evolve {}
   # Driver
   Specify Oxs_MinDriver [subst {
                basename [list $basename]
                evolver :evolve
                stopping_mxHxm 0.1
                checkpoint_interval 5
                stage_iteration_limit 1000
                mesh :mesh
                Ms :Mstot
                m0 :init
                projection_outputs {
                Mfx :mxout_f
                Mfy :myout_f
                Mfz :mzout_f
                Mrx :mxout_r
                Mry :myout_r
                Mrz :mzout_r
```

```
                }
    }]
# end the simulation
proc PoIn { Ms x y z } {
    if {$x<0 || $x>1 || $y<0 || $y>1} {return 0.0}
    return $Ms
}
proc RoundedCorners { Ms curad asprad x y z xspan yspan zspan} {
    set xoff [expr {$x*$xspan-$xspan+$curad*$asprad}]
    set yoff [expr {$y*$yspan-$yspan+$curad}]
    if {abs($xoff)*$xoff/$asprad/$asprad+abs($yoff)*$yoff>$curad*$curad} {return 0.0}
    set xoff [expr {$x*$xspan-$xspan+$curad*$asprad}]
    set yoff [expr {-$y*$yspan+$curad}]
    if {abs($xoff)*$xoff/$asprad/$asprad+abs($yoff)*$yoff>$curad*$curad} {return 0.0}
    set xoff [expr {-$x*$xspan+$curad*$asprad}]
    set yoff [expr {$y*$yspan-$yspan+$curad}]
    if {abs($xoff)*$xoff/$asprad/$asprad+abs($yoff)*$yoff>$curad*$curad} {return 0.0}
    set xoff [expr {-$x*$xspan+$curad*$asprad}]
    set yoff [expr {-$y*$yspan+$curad}]
    if {abs($xoff)*$xoff/$asprad/$asprad+abs($yoff)*$yoff>$curad*$curad} {return 0.0}
    if {$x<0 || $x>1 || $y<0 || $y>1} {return 0.0}
    return $Ms
}
proc Stripes {ws v1x v1y v1z v2x v2y v2z x y z xspan yspan zspan} {
        set xoff [expr {$x*$xspan}]
    set yoff [expr {($y-0.)*$yspan}]
    set dialin [expr {$xoff+$yoff+100*$ws}]
    set strdef [expr {fmod($dialin,$ws)/$ws}]
    if {$strdef<0.5} {return [list $v1x $v1y $v1z ]} else {return [list $v2x $v2y $v2z]}
}
proc StripesLim {ws v1x v1y v1z v2x v2y v2z x y z xspan yspan zspan} {
        if {$z<0 || $z>1} {return [list 0.0 0.0 0.0] }
        set xoff [expr {$x*$xspan}]
    set yoff [expr {($y-0.)*$yspan}]
    set dialin [expr {$xoff+$yoff+100*$ws}]
    set strdef [expr {fmod($dialin,$ws)/$ws}]
    if {$strdef<0.5} {return [list $v1x $v1y $v1z ]} else {return [list $v2x $v2y $v2z]}
}
# specifications for data outputs
Destination archive mmArchive
Schedule DataTable archive Stage 1
Destination datatab mmDataTable
Schedule DataTable datatab Stage 1
Destination graph mmGraph
Schedule DataTable graph Stage 1
#Schedule Oxs_MinDriver::Magnetization archive Stage 1
Destination display mmDisp
Schedule Oxs_MinDriver::Magnetization display Stage 1
```

[i] J. T. Heron, J. L. Bosse, Q. He, Y. Gao, M. Trassin, L. Ye, J. D. Clarkson, C. Wang, J. Liu, S. Salahuddin, D. C. Ralph, D. G. Schlom, J. Iniguez, B. D. Huey, and R. Ramesh, "Deterministic switching of ferromagnetism at room
temperature using an electric field", Nature v. 516, p. 370 (2014).
[ii] J. T. Heron, D. G. Schlom, and R. Ramesh, "Electric field control of magnetism using BiFeO3-based heterostructures", Applied Physics Reviews 1, 021303 (2014).
[iii] J. J. Wang, J. M. Hu, T. N. Yang, M. Feng, J. X. Zhang, L. Q. Chen, and C. W. Nan, "Effect of strain on voltage-controlled magnetism in BiFeO3-based heterostructures", Scientific Reports v. 4, 4553 (2014).